%% file: paper.tex
\documentclass[sigconf]{acmart}
\usepackage{balance,url,textcomp,multirow}

\usepackage{xcolor}
\usepackage{makecell}
\usepackage{graphicx}
\usepackage{algorithmic}
\usepackage{float,color}
\usepackage{xcolor}
\usepackage{gensymb}
\usepackage{geometry}
\usepackage{enumitem}

\newcommand{\squishlist}{
 \begin{list}{$\bullet$}
  { \setlength{\itemsep}{0pt}
     \setlength{\parsep}{3pt}
     \setlength{\topsep}{3pt}
     \setlength{\partopsep}{0pt}
     \setlength{\leftmargin}{1.5em}
     \setlength{\labelwidth}{1em}
     \setlength{\labelsep}{0.5em} } }

\newcommand{\squishlisttwo}{
 \begin{list}{$\bullet$}
  { \setlength{\itemsep}{0pt}
     \setlength{\parsep}{0pt}
    \setlength{\topsep}{0pt}
    \setlength{\partopsep}{0pt}
    \setlength{\leftmargin}{2em}
    \setlength{\labelwidth}{1.5em}
    \setlength{\labelsep}{0.5em} } }

\newcommand{\squishend}{
  \end{list}}
  
\usepackage[boxed,commentsnumbered,ruled,vlined,linesnumbered]{algorithm2e}

\SetCommentSty{mycommfont}

\SetKwInput{KwInput}{Input}                
\SetKwInput{KwOutput}{Output}              

\copyrightyear{2024}
\acmYear{2024}
\setcopyright{rightsretained}
\acmConference[EWSN '24]{International Conference on Embedded Wireless Systems and Networks}{December 10th-13th, 2024}{Abu Dhabi, UAE}
\acmConference[EWSN '24]{International Conference on Embedded Wireless Systems and Networks}{December 10th-13th, 2024}{Abu Dhabi, UAE}
\acmDOI{10.1145/3576842.XXXX}
\acmISBN{979-8-4007-xxxx}

\begin{document}

\title{I Still See You: Why Existing IoT Traffic Reshaping Fails}

\author{Su Wang, Keyang Yu, Qi Li, Dong Chen}
\affiliation{
  \institution{Colorado School of Mines}
  \city{Golden}
  \state{Colorado}
   \country{USA}
}
\email{{suwang, yukeyang, liqi, dongchen}@mines.edu}

\begin{abstract}
\input{abstract}
\end{abstract}

\ccsdesc[300]{Computing methodologies~Model development and analysis}
\ccsdesc[300]{Computing methodologies~Machine learning approaches}
\ccsdesc[300]{Computing methodologies~Neural networks}
\ccsdesc[300]{Computing methodologies~Classification and regression trees}

\keywords{IoT Privacy, Inference Attack, Image Processing, Deep Learning, Data Analytics, Smart Homes}

\maketitle


\section{Introduction}
\label{sec:introduction}
\input{introduction}

\section{Background and Related Work}
\label{sec:background}
\input{background}

\section{Challenges}
\label{sec:challenges}
\input{challenges}

\section{Design}
\label{sec:design}
\input{design}

\section{Implementation}
\label{sec:implementation}
\input{implementation}

\section{Experimental Evaluation}
\label{sec:evaluation}
\input{evaluation}

\section{Conclusion and Future Work}
\label{sec:conclusion}
\input{conclusion}

\bibliographystyle{ACM-Reference-Format}
\bibliography{paper}

\end{document}

%% file: abstract.tex
The Internet traffic data produced by the Internet of Things (IoT) devices are collected by Internet Service Providers (ISPs) and device manufacturers, and often shared with their third parties to maintain and enhance user services. Unfortunately, on-path adversaries could infer and fingerprint users' sensitive privacy information such as occupancy and user activities by analyzing these network traffic traces. While there's a growing body of literature on defending against this side-channel attack---malicious IoT traffic analytics (TA), there's currently no systematic method to compare and evaluate the comprehensiveness of these existing studies. To address this problem, we design a new low-cost, open-source system framework---IoT Traffic Exposure Monitoring Toolkit (ITEMTK) that enables people to comprehensively examine and validate prior attack models and their defending approaches. In particular, we also design a novel image-based attack capable of inferring sensitive user information, even when users employ the most robust preventative measures in their smart homes. Researchers could leverage our new image-based attack to systematize and understand the existing literature on IoT traffic analysis attacks and preventing studies. Our results show that current defending approaches are not sufficient to protect IoT device user privacy. IoT devices are significantly vulnerable to our new image-based user privacy inference attacks, posing a grave threat to IoT device user privacy. We also highlight potential future improvements to enhance the defending approaches. ITEMTK's flexibility allows other researchers for easy expansion by integrating new TA attack models and prevention methods to benchmark their future work.

%% file: introduction.tex
People are increasingly deploying the Internet of Things (IoT) devices to automate their smart homes. The number of IoT devices worldwide is forecast to almost double from 15.1 billion in 2020 to more than 29 billion IoT devices in 2030~\cite{Statista}. To maintain and enhance customer services, network traffic traces generated by these IoT devices is typically recorded by multiple on-path service providers and their third parties. These may include Internet Service Providers (ISPs), IoT device manufactures, cloud service providers, and their third parties. Verizon uses ``supercookies" to track their customers' Internet traffic activity, and AT\&T charges customers an extra \$29 per month if they would like to avoid ``the collection and monetization of their Internet browsing history for targeted ads," Mozilla told Congress~\cite{mozilla}. Recent research work~\cite{PrivacyGuard,Paros,apthorpe2019keeping} explained that ISPs like AT\&T, Comcast, and Verizon are selling personal traffic data without prior user consent~\cite{forbes}. Also, recent IoT privacy survey~\cite{leakage-new} shows that 72 out of 81 popular IoT devices in the U.S. and the U.K. are sharing data with third parties (e.g., Google, Amazon, Akamai) completely unrelated to original manufacturer and far beyond necessary device configuration and maintenance, including voice speakers, doorbells, thermostats, smart TVs, and streaming dongles, further exacerbating the situation. 

Meanwhile, significant recent research~\cite{Ding:2018:SID:3243734.3243865,park2014energy,cai2014systematic,chen2014combined,bovornkeeratiroj2020repel,wang2017walkie,wang2008dependent,shmatikov2006timing,apthorpe2019keeping,dyer2012peek,juarez2016toward} shows that launching traffic analytics (TA) attacks is surprisingly easy, since user activities highly correlate with simple time-series data statistical metrics. Thus, IoT device traffic rates alone have significant user privacy threats. To defend against these side-channel attacks, extensive prior research~\cite{park2014energy,dingledine2004tor,cai2014systematic,chen2014combined,bovornkeeratiroj2020repel,wang2017walkie,wang2008dependent,shmatikov2006timing,apthorpe2019keeping,dyer2012peek,raghavan2014modeling,hasan2008reconfigurable,rothmeier2020prediction,Paros,PrivacyGuard} proposed significant work to thwart privacy attacks on IoT traffic rates. Unfortunately, these prior approaches usually assumed different privacy threat models to design their approach and evaluated their work's performance using different datasets and different evaluation metrics. Despite the increasing volume of literature addressing the defense against these malicious IoT traffic analytics, there is currently a lack of a systematic method to compare and assess the comprehensiveness of these existing studies. Reproducing, benchmarking, and validating the efficiency and completeness of their results is impractical.

To address these problems, we design a new open-source system framework---\underline{\textbf{I}}oT
\underline{\textbf{T}}raffic \underline{\textbf{E}}xposure \underline{\textbf{M}}onitoring \underline{\textbf{T}}ool\underline{\textbf{k}}it (ITEMTK) that enables people to comprehensively examine and validate prior attack models and their defending approaches. In addition, ITEMTK provides a full stack performance evaluation for attack models and privacy preserving approaches. As we develop the system to get a thorough understanding of previous methods, we also discover that existing defense mechanisms fall short in safeguarding user privacy. Specifically, we've devised a novel image-based attack capable of inferring sensitive user information, even when users employ the robust preventative measures in their homes. Thus, smart home IoT devices are significantly vulnerable to our new image-based user privacy inference attacks, posing a grave threat to IoT device user privacy. By doing so, we make the following technical contributions.

\noindent {\textbf{Challenges}}. We explore and highlight the major challenges to design of ITEMTK, which encompass inconsistent attack models, datasets, and evaluation metrics, closed-source coding, as well as the uncertain encoding or representation of time-series to images.

\noindent {\bf ITEMTK Design}. We present the design of a new systematic framework---ITEMTK, which enables people to examine TA attacks and their defense approaches. ITEMTK has multiple components, including traffic data preprocessing, sophisticated TA attack defending, intelligent TA attacking, and full stack evaluation. In essence, ITEMTK initially utilizes a data collector to gather all publicly accessible IoT traffic datasets and preprocess them to prepare for the application of TA attack defense approaches. ITEMTK then applies the most recent TA attack defense approaches. By doing so, ITEMTK is securing smart homes using most recent user privacy masking approaches. ITEMTK then will launch a wide set of adversarial machine learning attacks, and our new image representation based fusion attack. Lastly, ITEMTK will perform a full stack benchmarking and evaluation on all the residual traffic rates. 

\noindent {\bf Implementation and Evaluation}. We implement ITEMTK in {\tt python} using widely available open-source frameworks. We implemented four different image representations, including Line Chart, Heat Map, Scatter Plot, and Gramian Angular Fields (GAF) images. Our evaluation results have shown that current defending approaches are not sufficient to protect IoT device user privacy. IoT devices are significantly vulnerable to our new image-based user privacy inference attacks, posing a grave threat to IoT user privacy.

\noindent {\bf Releasing Datasets and Code}. Our approaches to examine user sensitive information leakage through IoT traffic rates are quite general, and can be applied to address similar user privacy problems in other domains, such as medical IoT data analytics and smart grid smart meter data analytics. We released the ITEMTK framework and TA attack defending source code (excluding TA attack models) to the broad IoT research community on our website~\cite{itemtk}.

%% file: background.tex
\subsection{Background}

The research focuses on the ``edge'' of the Internet of Things (IoT)---namely, the interaction between the Internet and smart devices, i.e., smart homes. We assume smart homes use Wi-Fi gateways from the ISPs to connect to the public Internet. Smart homes may have routers, IoT Hubs or other similar devices that support multiple IoT devices to access to Internet. The smart home could employ numerous IoT devices including smart sensors, smart devices, and smart appliances. Many IoT devices are web-enabled with the ability to interact with cloud-based services, such as Google Home, Amazon Alexa, etc. These cloud services collect data from IoT devices and enable the remote control of these IoT devices. And the remote control could be performed by end users using their microphones, smart phone apps or automatically based on pre-defined user preferences.

\begin{figure}[t]
\begin{center}
\begin{tabular}{cc}       
\includegraphics[width=0.22\textwidth]{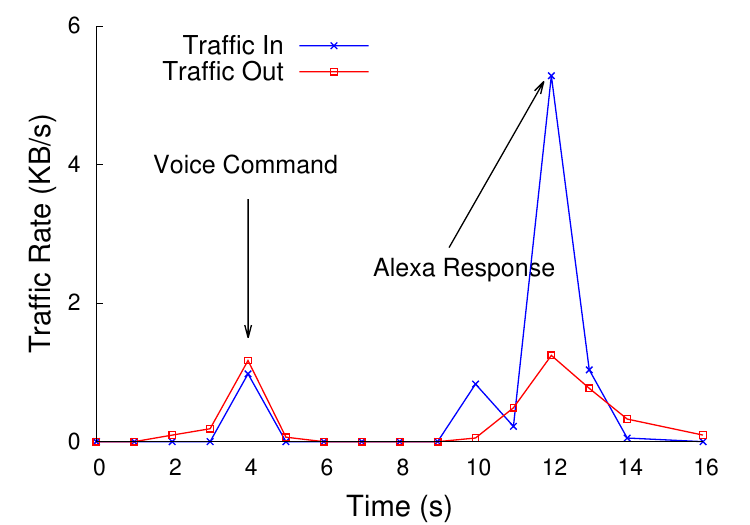} &
\includegraphics[width=0.22\textwidth]{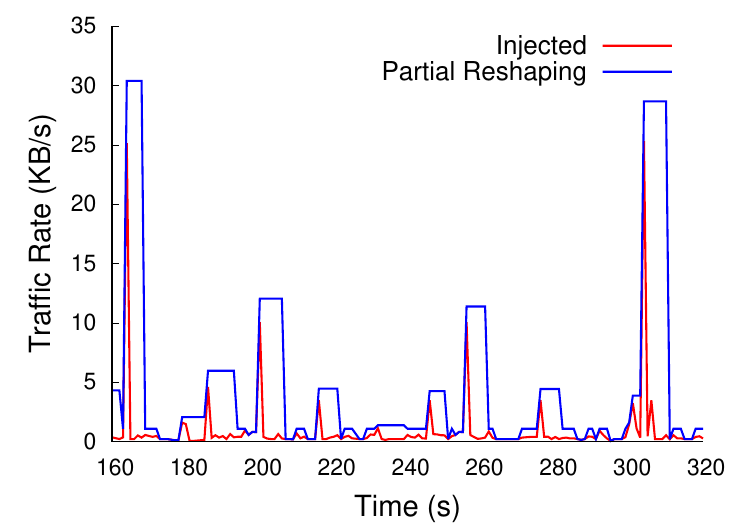}\\
(a) Amazon Alexa Dot & (b) Traffic Reshaping
\end{tabular}
\end{center}
\vspace{-0.1in}
\caption{The illustration of IoT traffic.}
\label{fig:traffic-reshaping}
\vspace{-0.2in}
\end{figure}

In modern smart homes, many IoT devices are user interaction intensive, e.g., voice assists, IoT smart plugins. Thus, as shown in Figure~\ref{fig:traffic-reshaping} (a), they may present bidirectional network traffic flows, including both outgoing and incoming network traffic flows. Figure~\ref{fig:traffic-reshaping} (a) shows Amazon Alex Dot's network traffic rate trace for 150 minutes. Device types and their associated in-home activities can be clearly identified using those traffic bursts (a.k.a, motifs or signatures in most recent work~\cite{PrivacyGuard,Paros}). To mitigate or prevent this user private information leakage from their IoT traffic rates, there are significant approaches~\cite{park2014energy,dingledine2004tor,cai2014systematic,chen2014combined,bovornkeeratiroj2020repel,wang2017walkie,wang2008dependent,shmatikov2006timing,apthorpe2019keeping,dyer2012peek,raghavan2014modeling,hasan2008reconfigurable,rothmeier2020prediction,Paros,PrivacyGuard} presented in the literature. The main goal of these research is to develop an effective algorithm that can ensure if an arbitrary single substitution in the IoT traffic rates is small enough, the adversaries can not infer accurate user in-home sensitive information. The broad idea for these approaches is to design some network traffic reshaping or padding algorithms to modify the spikes presented in their IoT traffic rates. Thus, attackers might become confused and struggle to distinguish genuine traffic motifs from artificial ones.

\subsection{Privacy Threat Model}

\begin{figure}[t]
\centering              
\includegraphics[width=0.5\textwidth]{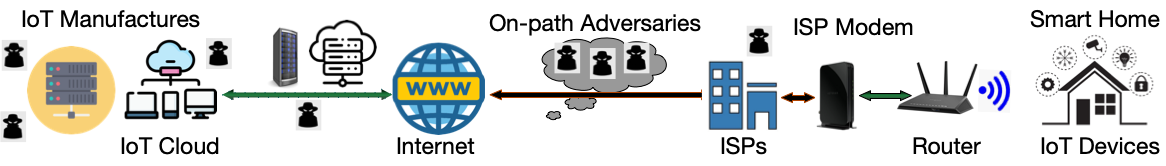}
\caption{Privacy Threat Model.}
\label{fig:threat-model}
\vspace{-0.2in}
\end{figure}

As shown in Figure~\ref{fig:threat-model}, in the prior side-channel attack research work, people are broadly concerned with the ability of external adversaries (e.g., ISPs, on-path network observers, IoT manufactures, and their third parties) to infer user sensitive activities from their IoT network traffic rate metadata. The network traffic rate metadata, including inbound/outbound traffic rates, network protocols, source, and destination IPs and package sizes, are accessible to many on-path external entities. These external adversaries can view smart home aggregated traffic data only after it has left the home local area networks (LAN). Note that, certain research works assumed attackers have more in-depth knowledge (e.g., DNS requests, MAC addresses) about their IoT network traffic. This kind of information typically is only available for internal adversaries.

And these potential adversaries may be incentives to infer user activities in smart homes where users do not want to share this privacy-sensitive information with them. We assume external adversaries can use advanced TA attack techniques, such as machine learning (ML), deep learning (DL), or Artificial Intelligence (AI) powered methods to infer certain types of the embedded user activity pattern information in the recorded traffic rates. Thus, inferring user activities in these homes is considered as an opposition to their users' privacy preferences. In particular, people are concerned with four user privacy inference attacks: 

\begin{itemize}[leftmargin=*]

\item \textit{Learning user occupancy.} This includes whether a home or building is occupied and when, and whether the home or building has multiple occupants; 

\item  \textit{Learning network traffic pattern information.} This includes whether a particular IoT device (e.g., Voice Assistant, Doorbell Camera, myQ-connected Smart Garage) is present in a home, and how much traffic and how often the home consumes on it monthly; 

\item \textit{Learning short-term user activities.} These user activities are inferred using IoT device activities and may include when users come and go, going to bed, waking up, watching TV, listening to music, playing online games, or having parties; 

\item \textit{Learning long-term user activities.} In addition, adversaries may also be interested in inferring more comprehensive and longer-term user activities, such as whether they have personal health conditions, whether they have night shift jobs or work at home, and whether they go on vacation on weekends or holidays.
\end{itemize}

\noindent{\bf Attack Scenario \#1}: To infer the type of IoT devices at a smart home, an external on-path adversary intends to collect real-time IoT network traffic traces and leverages ML/DL/AI-powered advanced data mining approaches to learn whether a particular IoT device (e.g., Amazon Alexa, Google Home, Ring Doorbell) is present in the home and how often this device is used daily or weekly. Then, the external attacker could launch cyberattacks on the specific device.

\noindent{\bf Attack Scenario \#2}: An external Internet on-path adversary (e.g., ISPs, IoT device manufacturers or their third-parties) is actively monitoring the IoT traffic traces for some target smart homes or buildings for some time. Then, the adversary may use their traffic analytics (TA) attack approaches to analyze traffic ``motifs'' and thus can learn indirect user private information (e.g., user short-term and long-term activities) that might be interesting for insurance companies, marketers, installers, or government agencies. 

\subsection{IoT Traffic Reshaping Roadmap}

There's a growing body of literature concerning against this malicious side-channel TA attack. We outlined traffic reshaping design alternatives that are most related to IoT traffic reshaping. In doing so, we review a wide range of the most recent sophisticated traffic reshaping-based prevention techniques~\cite{park2014energy,dingledine2004tor,cai2014systematic,chen2014combined,bovornkeeratiroj2020repel,wang2017walkie,wang2008dependent,shmatikov2006timing,apthorpe2019keeping,dyer2012peek,raghavan2014modeling,hasan2008reconfigurable,rothmeier2020prediction,Paros,PrivacyGuard} to thwart privacy attacks on IoT traffic rates. Unfortunately, numerous previous methods are not open-source or are not releasing their datasets. To understand their performance, we implemented three traffic reshaping approaches, including general pure traffic injection, general random traffic padding, and general hybrid traffic reshaping approaches in ITEMTK. 

\noindent{\bf Pure Traffic Injection (PTI)}. Prior work~\cite{park2014energy,cai2014systematic,dingledine2004tor} presented to inject artificial network traffic patterns to conceal genuine user network traffic patterns. Park et al.  discovered that traffic encryption alone is insufficient in preventing privacy invasions, as attackers can exploit vulnerabilities through traffic pattern analysis and statistical inference~\cite{park2014energy}. Additionally, they have created empirical models to statistically understand user behaviors based on transition data from wireless sensors. Subsequently, they inject cloaking network traffic patterns to obscure genuine traffic patterns. Cai et al. presented a defense---Tamarow against Tor website fingerprinting that can reshape traffic rate traces by controlling the size of the parameter to pad Internet traffic packets~\cite{cai2014systematic}. 

\noindent{\bf Random Traffic Padding (RTP)}. Recent research~\cite{apthorpe2019keeping,dyer2012peek,juarez2016toward} has introduced defending approaches based on random traffic padding. These methods aim to prevent external adversaries from reliably distinguishing genuine user-involved traffic patterns from ``fake'' ones. Dyer et al. proposed a buffered fixed-length obfuscator using random padding to prevent website fingerprinting attacks~\cite{dyer2012peek}. Juarez et al. suggested an adaptive padding approach that offers a sufficient level of security against website fingerprinting by matching gaps between traffic packets with a distribution of generic network traffic~\cite{juarez2016toward}. Similarly, Apthorpe et al. presented a stochastic traffic padding algorithm, which is deployable on edge gateways, middle boxes, or IoT hubs. This algorithm flattens traffic rates and injects fake traffic patterns that resemble genuine IoT traffic patterns~\cite{apthorpe2019keeping}.

\noindent{\bf Hybrid Traffic Reshaping (HTR)}. Prior work~\cite{chen2014combined,bovornkeeratiroj2020repel,shmatikov2006timing,wang2008dependent,raghavan2013coupled,raghavan2014modeling,hasan2008reconfigurable,rothmeier2020prediction,PrivacyGuard} presented hybrid reshaping techniques as a countermeasure against user privacy leakage from IoT traffic rates. Chen et al. proposed learning the ``noise'' injection rate through empirical analytics of IoT device activities~\cite{chen2014combined}. Similarly, Bovornkeeratiroj et al. proposed RepEL which employed an edge gateway (typically, a Raspberry PI-class node) to partially flatten traffic loads and randomly replay traffic loads to hide user occupancy information~\cite{bovornkeeratiroj2020repel}. Shmatikov et al. proposed adaptive padding algorithms to destroying timing ``fingerprints'' application traffic by enforcing inter-package intervals to match pre-defined probability mass functions~\cite{shmatikov2006timing}. Wang et al.~\cite{wang2008dependent} designed a traffic padding algorithm employing matched package schedules to prevent adversaries from pairing incoming and outgoing traffic. Significant work ~\cite{raghavan2013coupled,raghavan2014modeling,hasan2008reconfigurable,rothmeier2020prediction} proposed to model user activities using Markov Chain model. Keyang et al. proposed PrivacyGuard~\cite{PrivacyGuard} assumed the installation of an additional IoT hub to concurrently reshape both incoming and outgoing traffic. They also presented the design of PAROS~\cite{Paros}, which enables users to regain the control on reducing their privacy leakage. PAROS leverages traffic signature learning, hidden Markov model (HMM)~\cite{rabiner1986introduction}-based artificial traffic injection, and partial traffic padding to obfuscate user privacy. Table~\ref{table:reshaping_comparison_1} quantifies the effectiveness of the recent 13 defending approaches. We use $\epsilon$-security~\cite{park2014energy} to describe the probability of a traffic reshaping approach cannot prevent users from external adversarial inferring attacks. \emph{Lower values indicate a stronger guarantee in ITEMTK}. 

\noindent{\bf Observation}. We use $\epsilon$-security~\cite{park2014energy} to describe the probability of a traffic reshaping approach can not prevent smart home users from an external adversary's in-home activity inferring. Table~\ref{table:reshaping_comparison_1} shows that on average pure traffic injection, random traffic reshaping, and hybrid traffic reshaping approaches yield the $\epsilon$-security ranging from 3.4\% to 87.15\%, respectively. Unsurprisingly, pure traffic injection approach reports the highest $\epsilon$-security as of 87.15\%. This is mainly due to the fact that pure traffic injection approaches typically only injects and adjusts the shape of those artificial traffic patterns and thus not reshape any real ones already embedded in IoT traffic traces. However, these approaches did not always hide the genuine traffic patterns, in particular, during higher and lower traffic periods. Also, their traffic injection was built in a simulator which would require a device to host it to reshape traffic rates. The system overhead is not fully evaluated towards real deployments. Instead, hybrid traffic reshaping approaches report better $\epsilon$-security (e.g., PrivacyGuard~\cite{PrivacyGuard}, RepEL~\cite{bovornkeeratiroj2020repel}, PAROS~\cite{Paros}). The hybrid traffic reshaping approach strives to partially flatten both genuine and artificial traffic patterns. The random traffic padding approach typically employs a higher flattening threshold to pad IoT traffic patterns and considers bidirectional traffic padding for IoT devices like Amazon Alexa and Google Home. Due to the nature of random traffic injection, these approaches may still allow external attackers to identify the injected ``fake'' signatures, and thus infer genuine user activities. These approaches typically assumed the installation of either a simulator  (on a computer) or edge gateway (typically, a Raspberry PI-class node) to enable their defending approaches, except the recent PAROS~\cite{Paros} work. For the same reason, Tamarow~\cite{cai2014systematic} reports the maximum traffic overhead as of 199\% additional overhead per device per day. The general implementation of pure traffic injection approaches yield additional overhead as $\sim$97\%. While, the general random traffic padding approach yields additional overhead as $\sim$165.9\%.

\begin{table}[t!]
\small
\begin{center}
\begin{tabular}{||l||l||c|c|c||}
\hline
\textbf{} & \textbf{Defenses} & \textbf{\makecell{Additional\\ Hardware}} &  \textbf{\makecell{Security\\ ($\epsilon$)}} & \textbf{\makecell{Additional\\ Overhead}}\\ \hline
{PTI} & {General} & Yes &87.15\% &97\%  \\ \hline
{PTI} & {Tor~\cite{dingledine2004tor}} & Yes   &77.5\% &25\% \\  \hline
{PTI} & {BUFLO~\cite{cai2014systematic}} & Yes &41.5\% &199\%  \\ \hline
{RTP} & {General} & Yes  &54.33\% & 165.9\% \\  \hline
{RTP}& {Tamarow~\cite{cai2014systematic}} & Yes  &3.4\% & 199\%  \\ \hline
{RTP} &{EPIC~\cite{liu2018epic}} & Yes &31.0\% &76.3\%  \\ \hline
{RTP} & {WTF-PAD~\cite{juarez2016toward}}   & Yes  &26\% & 54\% \\  \hline
{HTR} & {General} & Yes &72.6\%  &103.7\%  \\ \hline
{HTR}& {RepEL~\cite{bovornkeeratiroj2020repel}} & Yes   &33\% & 100\% \\  \hline
{HTR}& {PrivacyGuard} & Yes  & 14.2\% & 66.7\%  \\ \hline
{HTR}& {PAROS~\cite{10230103}} & No  & 15.3\% & 42.4\%  \\ \hline
{HTR}& {Energy~\cite{park2014energy}}    &Yes  &42.5\% & 40\% \\  \hline
{HTR}& {RepEL~\cite{bovornkeeratiroj2020repel}}    &Yes  & 33\% & 100\% \\  \hline
\end{tabular}
\end{center}
\caption{The comparison of 13 recent defending approaches when encountering ML/DL enabled inference attacks.}
\label{table:reshaping_comparison_1}
\vspace{-0.3in}
\end{table}

\subsection{Summary and Insight}
\label{sec:insight}

Although recent research proposes significant work to thwart privacy TA attacks on IoT traffic rates, there's currently no systematic method to compare and evaluate the comprehensiveness of these existing studies. Unfortunately, different prior approaches assumed different privacy threat models and evaluated their work's performance using different datasets and different evaluation metrics. It is hard to reproduce, benchmark and validate their results across different approaches. Also, full stack evaluation (e.g., energy consumption, memory storage, CPU utilization) towards read-world deployments is missing. Many existing approaches are proposed, implemented and evaluated on a simulator which is potentially installed on a host device (e.g., desktop, middle box, edge gateway, WiFi access point). Although these approaches indicate that it is feasible to mask user sensitive information embedded in their network traffic rates with some system overhead, we are still pondering two questions regarding valuable existing work.

\begin{itemize}[leftmargin=*]
    \item Is it possible to extract user sensitive information from residual traffic rates even after deploying the most effective defenses against TA attacks?
    \item How effective are existing defense approaches when against TA attacks across different granularities of network traffic rates?
\end{itemize}

\begin{figure}[t!]
\begin{center}
\includegraphics[width=0.4\textwidth]{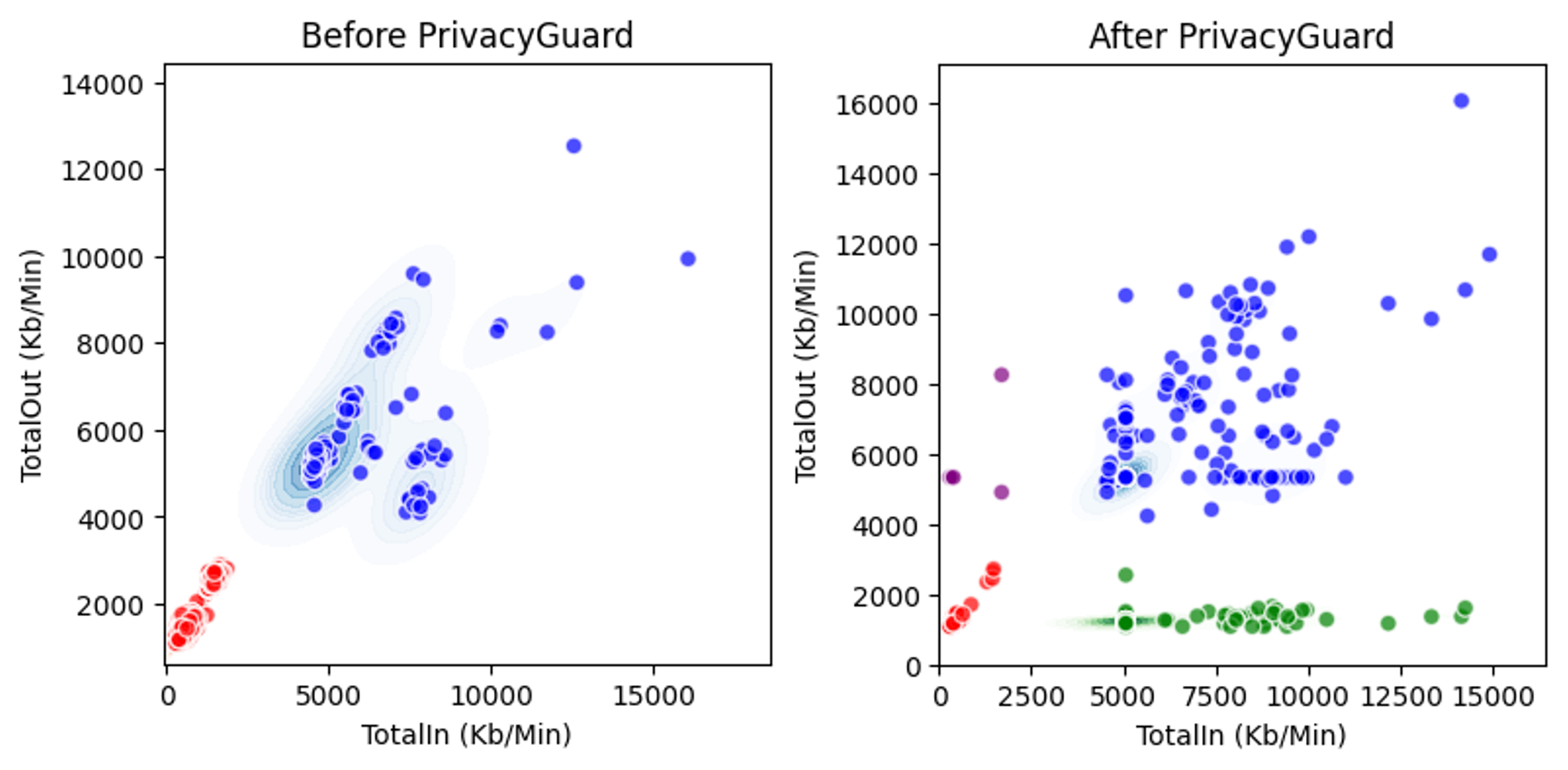}
\end{center}
\vspace{-0.3cm}
\caption{The residual traffic patterns (in red and blue colors) after applying most recent TA defense work.}
\label{fig:summary}
\vspace{-0.2in}
\end{figure}

Regarding the first question, interestingly, as shown in Figure~\ref{fig:summary}, we can still observe some ``patterns'' from residual network traffic rates in their image represented format after applying the recent TA defense approaches~\cite{PrivacyGuard,Paros}. This motivates us to further look into the problem and design a new image-based attack. We first represent the residual network traffic rate traces using different image formats and then design and implement a deep learning enabled fusion model to process and classify each data source to infer use in-home activities. As shown in our Section~\ref{sec:evaluation}, we find that current TA defending approaches are significantly susceptible to our new image-based attacks. For the second question, new practical systematized approach that can thoroughly examine and benchmark different attacking and defending approaches using same datasets (at different granularity) and same evaluation metrics are necessary. These valuable insights guide the development of our proposed open-source system framework---ITEMTK and also a new image-based TA attack in our design.

%% file: challenges.tex
In this section, we outlined the key challenges we met when designing, implementing, and evaluating our new system---ITEMTK.

\noindent{\bf Inconsistent Privacy Threat Models}. When designing ITEMTK, the first challenge is the definition and setup of user privacy threat model. This gives rise to several questions: Who are the adversaries targeting smart home IoT devices, where are they located, and what level of prior knowledge do they possess? To tackle this issue, we advocate for the development of the strictest privacy threat model. We do not naturally trust in IoT manufactures and their cloud service providers. We design ITEMTK from smart home user privacy guarantee perspectives. Additionally, we refrain from assuming in-depth prior knowledge in potential TA attacks. Instead, we propose evaluating defensive approaches against adaptive adversaries. This implies that adversaries may acquire some knowledge after monitoring a smart home over an extended period.

\noindent{\bf Inconsistent Attack Models}. Prior TA attack defense approaches use different attack models to evaluate the performance in terms of correctness, efficiency, and certain overhead of their presented algorithms and mechanisms. This makes it very hard to compare and validate the performance of different approaches. To overcome this challenge, we implement a wide set of attack models based on prior TA attack defense work~\cite{park2014energy,dingledine2004tor,cai2014systematic,chen2014combined,bovornkeeratiroj2020repel,wang2017walkie,wang2008dependent,shmatikov2006timing,apthorpe2019keeping,dyer2012peek,raghavan2014modeling,hasan2008reconfigurable,rothmeier2020prediction,Paros,PrivacyGuard}. In addition, we implement a set of ML and DL powered sophisticated adversarial attack models, which may better describe the capabilities of modern advanced on-path adversaries. 

\noindent{\bf Inconsistent Dataset and Evaluation Metrics}. Existing TA attack defense work often used different datasets and evaluation metrics to benchmark their approaches' performance. To overcome this challenge, we use the same big datasets to evaluate different TA attack defense approaches. In addition, we implement a wide set of evaluation metrics, such as F1 score, Matthews Correlation Coefficient (MCC), Precision, Recall, and weighted-average, to evaluate different approaches from the same benchmark perspectives. This will enable ITEMTK to fairly review prior TA defense works.

\noindent{\bf ``Closed'' Source Code and Evaluation Dataset}. The next challenge when we integrating all the prior TA attack models and defenses into our new system---ITEMTK is that we could not directly get the source code and their evaluation data from prior works. To overcome this problem, we went through their presented Pseudo-code and algorithms and then implement general approaches to benchmark their TA defense approaches' performance. We release these general approaches along with our system framework---ITEMTK. Note that, it is quite easy for researchers and other users to add new TA defense approaches into ITEMTK.

\noindent{\bf Time-series Data Representation}. As we discussed in Section~\ref{sec:insight}, prior attack models often focus on only one granularity of traffic rates to perform TA attacks and thus may still leave residual time-series traffic ``patterns'' that have embedded genuine use in-home activities. To address this issue, we propose to convert residual time-series traffic rates into different image representations, including Scatter Plot, Heat Map, Line Chart, and Gramian angular fields (GAF) images. The different image presentations could reserve the features presented in different granularities of traffic rates. We also need to handle input data alignment and processing acceleration issues for our image fusion based TA attack. We will discuss more details in Section~\ref{sec:design}. These challenges are well addressed in ITEMTK.

%% file: design.tex
\subsection{System Design}

While there's a growing body of literature on defending against this side-channel attack---malicious IoT traffic analytics, there's currently no systematic method to compare and evaluate the comprehensiveness of these existing studies. To address this problem, we design and implement a new framework---ITEMTK. In addition, ITEMTK also provides a full stack performance evaluation for TA attack models and their defense approaches. As we develop the system to get a thorough understanding of previous methods, we also find that existing defense mechanisms can not fully mask smart home user privacy. In particular, we also present a novel image-based attack capable of inferring sensitive user information, even when users employ the most robust TA attack preventative measures in their smart homes. Thus, smart home IoT devices are significantly vulnerable to our new image-based user privacy inference attacks, posing a grave threat to IoT device user privacy.

\begin{figure}[!t]
\includegraphics[width=0.42\textwidth]{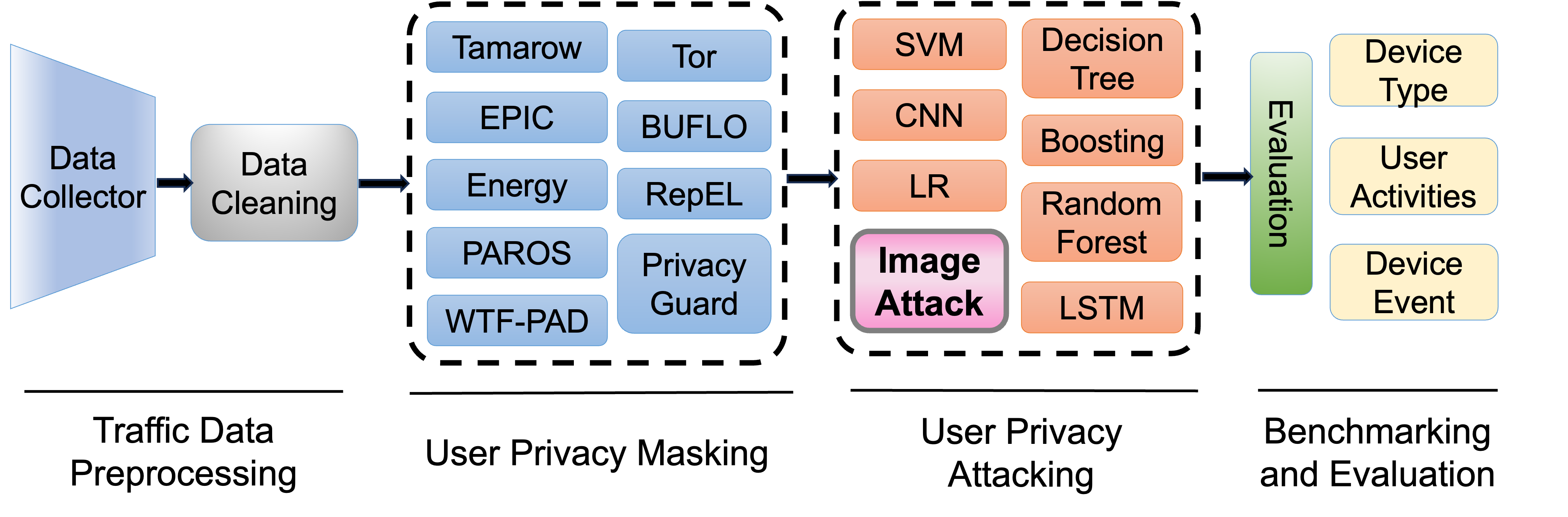}
\caption{The proposed structure of ITEMTK framework.}
\label{fig:itemtk-pipeline}
\vspace{-0.2in}
\end{figure}

\noindent{\textbf{System Operational Pipeline}}. Figure~\ref{fig:itemtk-pipeline} shows the system operational pipeline of our new framework---ITEMTK, which could enable people to comprehensively examine and validate prior TA attack models and their defense approaches. The whole pipeline of ITEMTK has multiple steps, including traffic data preprocessing, sophisticated TA attack defending, intelligent TA attacking, and full stack evaluation. In essence, ITEMTK initially utilizes a data collector to gather all publicly accessible IoT traffic datasets and then ``clean'' them to prepare for the application of TA attack defense approaches. ITEMTK then will apply the most recent TA attack defense approaches on the processed IoT network traffic rate data. By doing so, ITEMTK is securing smart homes using most recent user privacy masking approaches. Subsequently, ITEMTK will launch a wide set of ML/DL-powered adversarial attack models, and our new image representation based attack model. Lastly, ITEMTK will perform a full stack benchmarking and evaluation over all the residual traffic rate data. ITEMTK could compare the correctness, efficiency, and overhead against different TA defenses.

\subsection{Traffic Data Preprocessing}

The inputs of ITEMTK are the whole-home network traffic rates. The aggregated traffic rate data could be observed by the on-path adversaries. Next, ITEMTK will perform statistical analytics on the converted common data file to ensure its correctness and effectiveness. Then, ITEMTK will preprocess the data into NumPy arrays and split the whole dataset into training, testing and validation datasets. Also, ITEMTK leverages KNNImputer to fill in missing values using k-Nearest Neighbors approach. By default, a euclidean distance metric that supports missing values, is used to find the nearest neighbors. Each missing traffic rate is imputed using values from n\_neighbors nearest neighbors that have a value for the traffic rate. The traffic rates of the neighbors are averaged uniformly or weighted by distance to each neighbor. Our traffic spikes or motifs are located based on local maximum points. With optimal threshold and sliding window size, the total traffic rates can be extracted to independent motifs. The duration for each motif is the length of time that the traffic volume is continuously higher than given threshold and no longer than the sliding window size. Notably, sliding window size only limits the maximum duration of a traffic rate motif. Given a sliding window size $n$, threshold $T$, and local maximum point $x_p$, a traffic rate motif that has maximum duration (equals to $2n+1$) can be expressed as $x_{p-n}, x_{p-n+1}, ..., x_{p-1}, x_p, x_{p+1}, ..., x_{p+n-1}, x_{p+n}$. The features include {\textbf{Range}} $R$, {\textbf{Mean}} $\mu$, {\textbf{Variance}} $\sigma^2$, {\textbf{Standard Deviation}} $\sigma$, and: 

\noindent{\bf Area}: The area of the region bounded by the traffic volume graph and also the given threshold within the sliding window. 
\begin{equation}
    A=\int_{p-n}^{p+n}{(x-T)}dx
\end{equation}

\noindent{\bf Skewness}: The asymmetry of the selected ``spike'' about its local maximum traffic rate. 
\begin{equation}
    \widetilde{\mu_3}=\frac{1}{2n}\sum_{i=p-n}^{p+n}(\frac{x_i-\mu}{\sigma})^3
\end{equation}

\noindent{\bf Coefficient of Variation}: The standardized measurement of dispersion. 
\begin{equation}
    CV=\frac{\sigma}{\mu}
\end{equation}

\subsection{Sophisticated TA Attack Defending}

Once the traffic rates are prepossessed, ITEMTK will apply the prior TA attack defending approaches (e.g., PAROS~\cite{Paros}, PrivacyGuard~\cite{PrivacyGuard}) on the ``clean'' IoT network traffic rates. By doing this, ITEMTK will ``arm'' smart home IoT devices using the recent TA attack defense mechanisms. Within ITEMTK framework, we implemented three different traffic reshaping approaches, including general pure traffic injection, general random traffic padding, and general hybrid traffic reshaping approaches in our proposed framework. As we discussed in Section~\ref{sec:background}, this categorization is based on the user privacy guarantee level established in previous works in the literature, as well as their associated traffic overhead. 

Currently, ITEMTK has already included these three different TA attack preventing approaches in design. Researcher and other users can directly use ITEMTK to benchmark their new TA attack models. We make concerted efforts to comprehend the related work, even though it is frequently proprietary. We undertake the design and implementation of generalized versions of these closed-source algorithms. For pure traffic injection approach, we are motivated by prior work~\cite{park2014energy,cai2014systematic,liu2018epic}. To design the general random traffic injection approach, we use the ideas from prior TA defense work~\cite{park2014energy,cai2014systematic,dingledine2004tor}. Regarding the design of hybrid traffic reshaping approach, we summarize the insight from prior  work~\cite{chen2014combined,bovornkeeratiroj2020repel,shmatikov2006timing,wang2008dependent,raghavan2013coupled,raghavan2014modeling,hasan2008reconfigurable,rothmeier2020prediction,PrivacyGuard}. We also integrate two open-source TA defense work into our framework---ITEMTK. Specially, when building and integrating these approaches, we use the same input traffic rate traces and the same traffic rate features for their learning models. This approach aids us in establishing a fair benchmarking environment. Note that, the way ITEMTK integrating new TA defense approaches is ``pluggable'', which means users could always add new approaches to ITEMTK to benchmark and compare against other approaches. 

\subsection{Intelligent Traffic Analytics (TA) Attacking}

\noindent{\textbf{ML- and DL-powered TA Attacks}}. We then focus on selecting the optimal ML model that can achieve the best accuracy to infer user activity. We investigate the most widely used ML classifiers in prior TA related work, including Logistic Regression, Support Vector Machines (SVMs), Random Forest, Decision Tree, Naive Bayes, Nearest Neighbors, Gaussian Process Regression. Specially, we also benchmark different kernels for SVMs, including linear, linear passive-aggressive, linear ridge, polynomial with 1$\sim$10 degrees, and radial basis function (RBF). The inputs of these ML models are principal features identified on significant traffic rate motifs in~\cite{Paros,PrivacyGuard}, including the duration, mean, maximum and minimum values, standard deviations, range, Skewness, variation coefficient, kurtosis, and area under the curve (AUC). We then leverage Principal Component Analysis (PCA)~\cite{PCA} to analyze principle features from network traffic rate traces. The goal is not only understanding the weighted importance of different features but also performing dimension reduction to save the model training time. The data transformation process from 10-dimensional space into a low-dimensional space will ensure low-dimensional representation retains the meaningful ``patterns'' in the original network traffic rates.

In addition, we also design a convolutional neural networks (CNNs)-based deep learning TA approach to infer user in-home activities from IoT network traffic rate traces. Our CNNs architecture is inspired by the most notable prior CNNs research---VGGnet~\cite{vgg}. The CNNs architecture is comprised of input, convolutional layers (ReLU), max pooling, fully-connected layers (with and without ReLU) and output. In addition, two fully-connected layers with ReLU and another fully-connected layer (without ReLU) are added to process the outputs. As shown in prior work~\cite{PrivacyGuard}, the granularity of traffic rates also significantly impacts the performance of the selected features. This is mainly due to fact that some features (e.g., AUC, duration) could become hidden and thus harder to detect on coarser granularity traffic rates. To fully evaluate this effect, ITEMTK will perform TA attacks on different granularities on traffic rate motifs. Table~\ref{table:attack-models-all} shows the attack performance comparison of the ML- and DL-powered TA attacks. We find that the original MCC dropped to $\sim$0.3 after applying the most effective defense approaches (e.g., PrivacyGuard~\cite{PrivacyGuard}, PAROS~\cite{Paros}, Traffic Padding~\cite{apthorpe2019keeping}). We use ITEMTK to validate that these recent TA defense approaches can effectively prevent a wide set of ML/DL-powered TA attacks.  

\begin{figure}[t!]
\begin{center}
\includegraphics[width=0.35\textwidth]{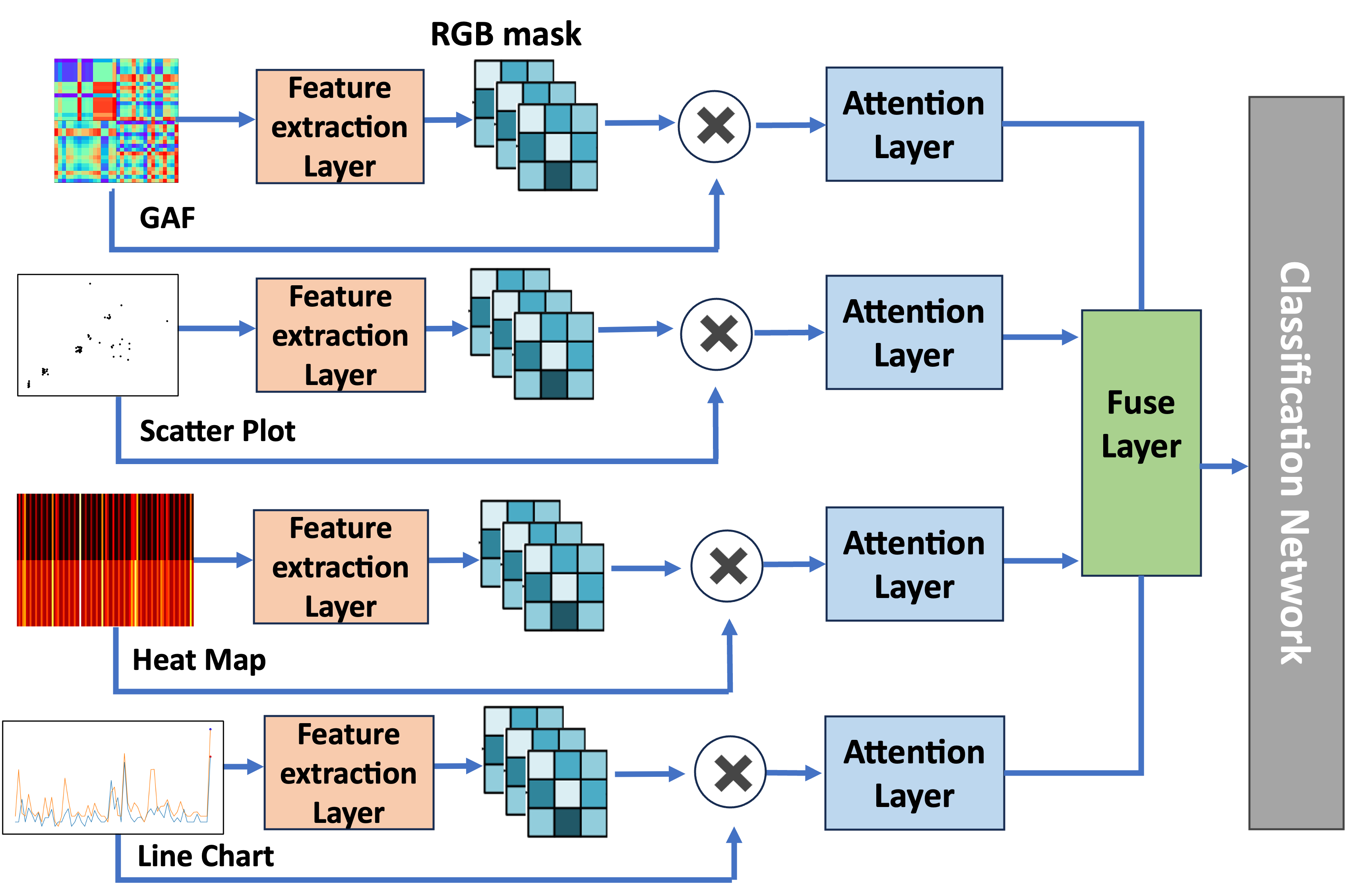}
\end{center}
\vspace{-0.3cm}
\caption{The system structure of our image-based TA attack.}
\label{fig:fusion}
\vspace{-0.2in}
\end{figure}

\noindent{\textbf{Novel Image Fusion based Smart Attacks}}. However, just like we discussed in Section~\ref{sec:background}, we discover that residual patterns persist in the traffic rates despite the application of the most recent TA attack defense approaches. To explore and benchmark this privacy threat, we design a new image presentation based TA attack, which coverts network traffic motifs from time-series format into multiple image representation formats. Then, our new attack will leverage a new image fusion model to detect different IoT devices and infer their associated user activities. Our fusion model can be used to classify various devices and their associated user activities. 

Data alignment and processing acceleration present significant challenges when building our model. It is essential to make all images into a consistent data format before fusion. In a fusion network, synchronizing the timing data across various image representations is crucial. The first step involves converting all time-series traffic rates into their visual formats. Considering that a single image representation may not be able to adequately represent the complexity of time series data, we employ a versatile fusion method that enables the selective integration of different image representations to observe more comprehensive features from the residual time-series traffic rate data. Specifically, we use the time and the in/out traffic rates to generate four types of visual representations, including Line Chart, Heat Map, Scatter Plot, and Gramian Angular Fields (GAF) image~\cite{barra2020deep} to further explore the residual traffic patterns in IoT network traffic rate traces. The GAF image representation in our ITEMTK design is motivated by another recent research work~\cite{barra2020deep} focusing on time series-to-image encoding for financial forecasting. Our insight regarding these time-series to image representation is that there are still significant genuine user activity embedded information in the residual traffic rate data processed by TA defense algorithms or mechanisms. In addition, image representation (e.g., GAF images) could represent and capture those hidden ``patterns'' represented in the residual traffic rate data at different granularities, which is missing in current TA attack and defense literature. 

\begin{figure}[t!]
\begin{center}
\includegraphics[width=0.3\textwidth]{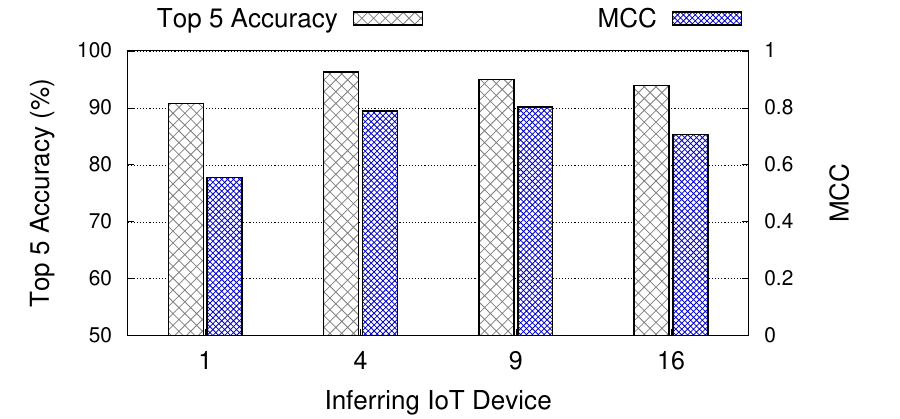}
\includegraphics[width=0.3\textwidth]{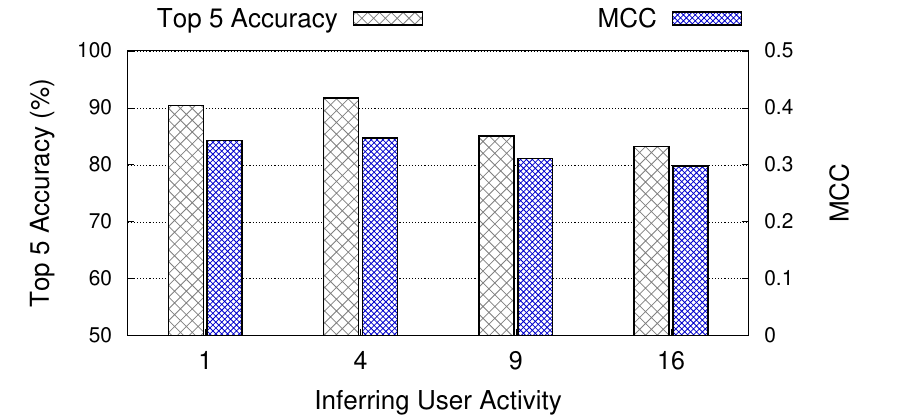}
\end{center}
\vspace{-0.3cm}
\caption{The GAF image representation when inferring IoT device events (top) and user activities (bottom).}
\label{fig:GAF-all}
\vspace{-0.2in}
\end{figure}

We then build a network for classification of user activities. Figure~\ref{fig:fusion} illustrates the structure of our image fusion network. The network is comprised of two main segments, including the fusion component and the classification component. The fusion component could integrate the visual represented traffic rate data, and also leverage 2D convolutional layers to extract spatial features from distinct perspectives of different image presentation. As shown in Figure~\ref{fig:GAF-all}, to understand how many images at different granularities we should include in our GAF image representation, we benchmark different configurations. The goal is to find the optimal number of sub-graphs that could collaboratively capture the most residual and sensitive information for user activity detection. We find that when GAF representation keeps four different granularities could observe the optimal use activities detection accuracy in terms of Top-5 accuracy (Y1 axis) and Matthews Correlation Coefficient (MCC~\cite{mcc}, on Y2 axis). The user activity inferring accuracy drops after $GAF\_Num = 4$ mainly because user activity is highly correlated with higher traffic spikes, which can be smoothed out or hidden as $GAF\_Num$ increases. Thus, ITEMTK uses four images based structure to convert time-series traffic rates into GAF images.

Given the feature sparsity presented in different image formats, we then apply targeted RGB masks on channels, further enhancing the our network's focus on key image areas. The processed images will be normalized and passed through activation functions to refine our learned features. Subsequently, an attention module is used to dynamically weigh the significance of the features in both pre- and post- concatenation processes. The classification component will use the fused features to perform the final user activity inference classification. The structure of our fusion model is also designed to be adaptable. Researchers and other users could easily integrate other baseline classification methodologies into our ITEMTK system framework. Note that, the visual image representations that ITEMTK uses is orthogonal to the other aspects of the technique and is thus ``pluggable,'' such that we could use other image representation approaches to perform time-series traffic rate data to image data encoding operations here. We will include new becoming online image representation technique in our future work.

\begin{figure}[t!]
\centering              
\begin{tabular}{cc}       
\includegraphics[width=0.21\textwidth]{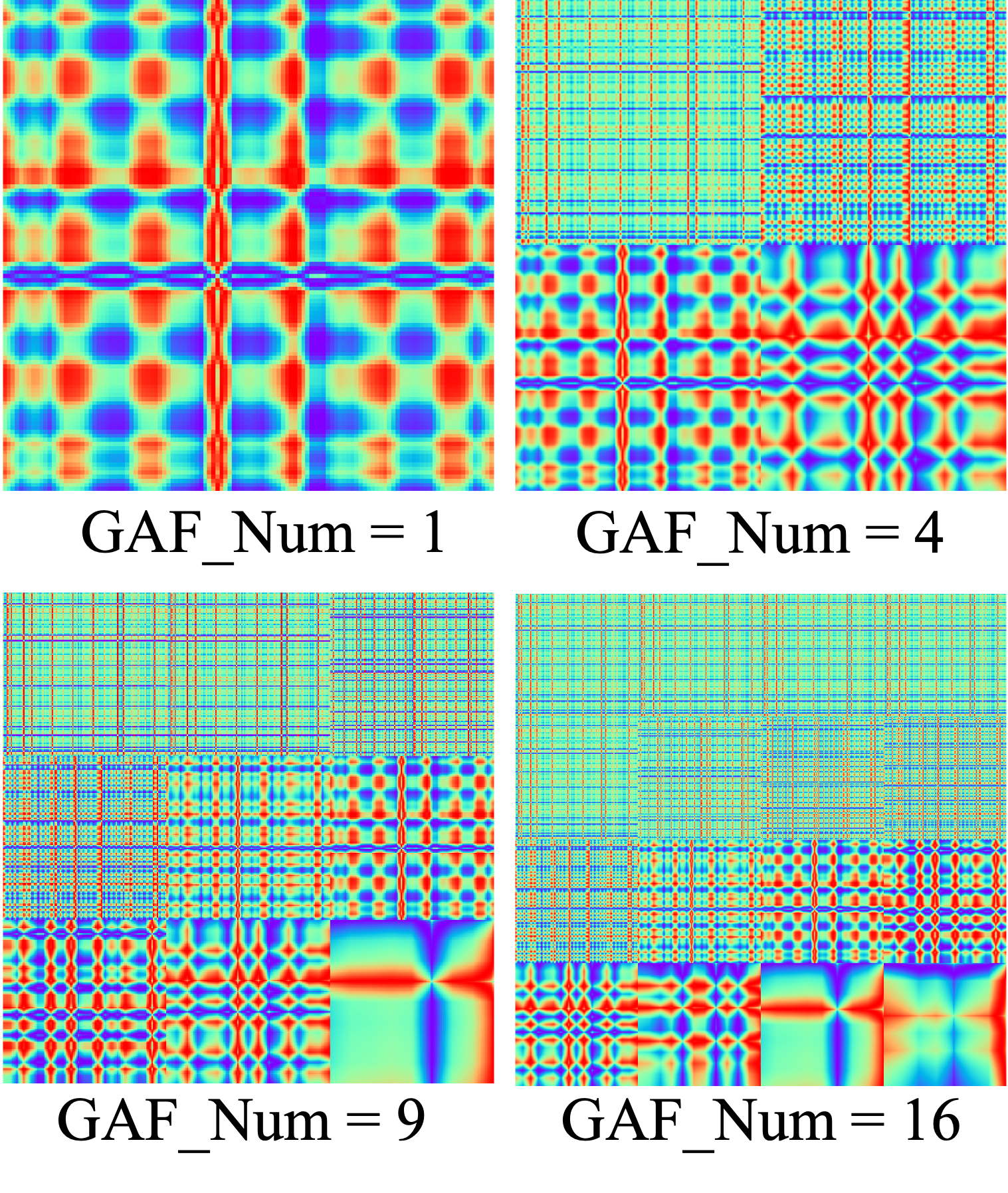} &
\includegraphics[width=0.21\textwidth]{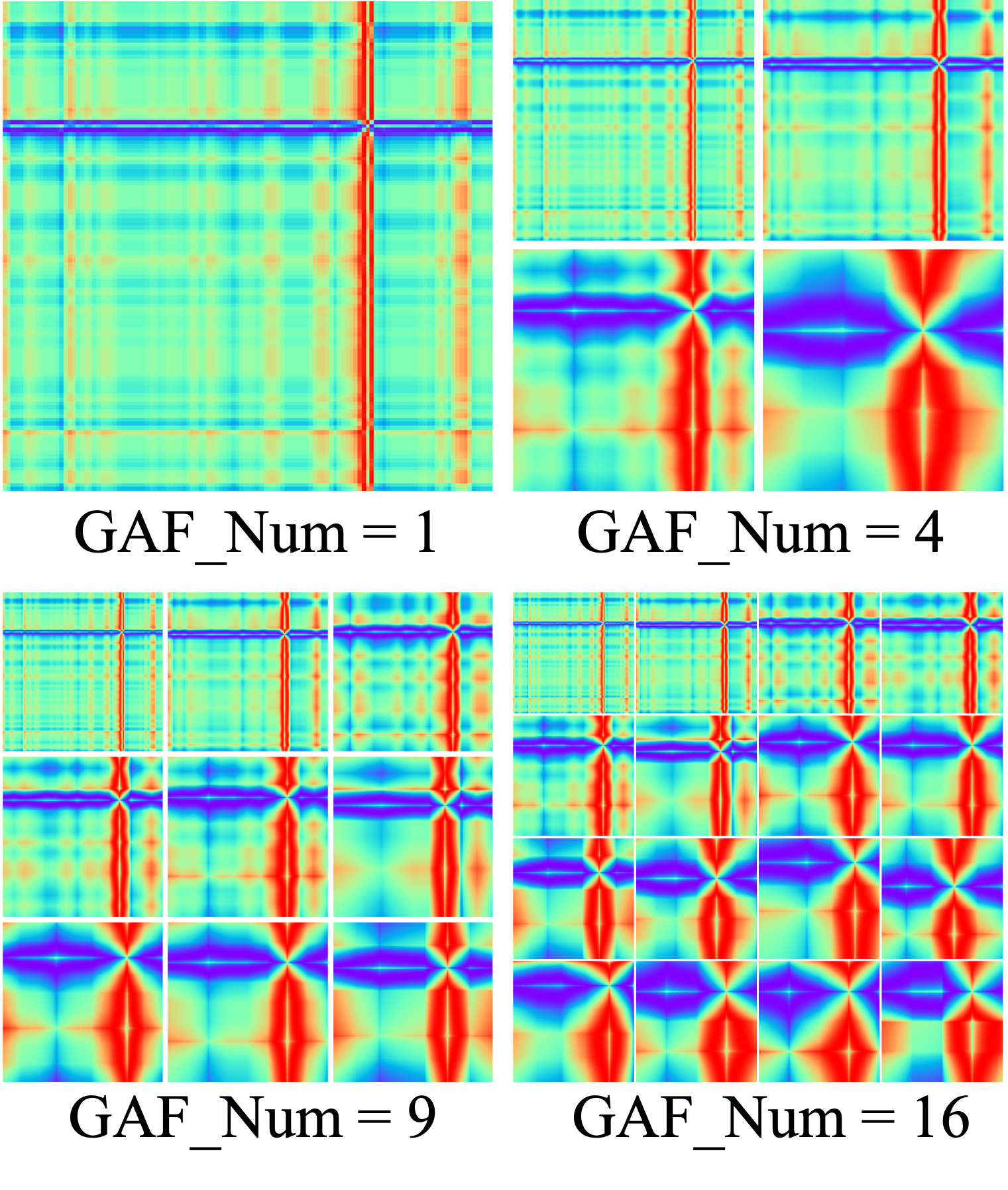}\\
(a) IoT device events & (b) user activities
\end{tabular}
\vspace{-0.2cm}
\caption{The illustration of our GAF representation when varying the amount of sub-graphs in each GAF image.}
\vspace{-0.2in}
\label{fig:GAF-number}
\end{figure}

\subsection{Full Stack Evaluation}
\label{sec:full-stack-evaluation}

The last component of ITEMTK framework is providing the full stack evaluation of the TA defense approaches, which is missing many prior works in the literature. ITEMTK provides a full stack TA attacks and their defense approaches related evaluations. In essence, ITEMTK has the capability to assess the effectiveness of privacy-preserving measures through various metrics, including F1 score, Matthews Correlation Coefficient (MCC)~\cite{mcc}, Precision, Recall, and Adversary Confidence (AC)~\cite{PrivacyGuard}. Additionally, ITEMTK provides support for assessing TA attacks launched by adaptive adversaries who might acquire extensive knowledge about a smart home through prolonged monitoring. The inference evaluation includes inferring IoT device types, detecting IoT device events, and learning their associated user activities. Eventually, ITEMTK also examines TA defense approaches in terms of their CPU utilization, RAM, ROM, Network Bandwidth, and I/O when they are protecting smart home IoT devices. That being said, ITEMTK can provide practical TA defense evaluations towards real world deployments.

%% file: implementation.tex
We implement ITEMTK in {\tt python} using widely available open-source frameworks, including Pandas~\cite{mckinney2011pandas}, Scikit-learn~\cite{pedregosa2011scikit} and PyCUDA~\cite{klockner2012pycuda}. ITEMTK takes a home's network traffic rate traces as input and applies most significant TA attack prevention techniques outlined in Section~\ref{sec:design}. To address missing data points issue, we leverage KNNImputer from Scikit-learn~\cite{pedregosa2011scikit} to implement the process to add those missing data points. We use the Scikit-learn~\cite{pedregosa2011scikit} library in {\tt python} to build our ML and DL based TA attack models. For CNNs-based attack approach, we implement the model based on the recent framework---VGGnet~\cite{vgg}. Regarding the image based attacks, we implement the image fusion network using torch2.0.1+cu118~\cite{paszke2019pytorch}. We have implemented four image representations, including Line Chart, Heat Map, Scatter Plot, and Gramian Angular Fields (GAF) image~\cite{barra2020deep}. Then, we leverage YOLOv5~\cite{zhu2021tph} to train our image fusion based detection models. The training parameters are set to 100 epochs with a batch size of 64. The image size for our models is 224. Our training process uses eight worker threads. The learning rate was set as of 0.001, with a weight decay of $5e^{-05}$. We implement label smoothing at a rate of 0.1. We followed the standards to assign 70\% to the training set, 15\% to the validation set, and the remaining 15\% to the testing set. We then implement pure traffic injection, random traffic padding, and hybrid traffic reshaping approaches in {\tt python}. For the recent defending approaches~\cite{Paros,PrivacyGuard}, we download their open-source codes and integrate them into our ITEMTK framework. We schedule batch jobs on GPU servers to compare F1 score, MCC, Precision, and Recall of different TA attack preventing approaches using CUDA. The server has resources: 1) CPU: 2x Intel(R) Xeon(R) CPU E5-2620 v4 @ 2.10GHz, 2) GPU: nVidia TITAN X (Pascal) (x8), 3) RAM: 128GB, 4) OS: Linux CentOS 7.

%% file: evaluation.tex
\subsection{Datasets}

\noindent{\bf Dataset: UNSW}. We chose the most widely used and publicly-available dataset---UNSW to benchmark all different approaches of ITEMTK. We downloaded the publicly-available IoT traffic traces from UNSW Sydney~\cite{sivanathan2018classifying} that includes second level traffic traces of 22 IoT devices for 20.5 days. We then process the IoT traffic metadata to traffic rates and also label all their associated user activities.

\noindent{\bf Traffic Rate Preprocessing}. To learn the effect of traffic rate granularity on user privacy preserving degree, we processed traffic rate traces of the above-mentioned datasets into different granularities, such as one second, one minute, three minutes, five minutes, and ten minutes. By default, traffic granularity is set as of one second. 

\noindent{\bf Traffic Rate Motif Extraction}. We also apply a sliding window as the size of 60 seconds to further process the traffic rate traces. The key insight is that numerous IoT devices exhibit bidirectional traffic flows with either the IoT manufacturer or their remote servers. Furthermore, a significant portion of these traffic flows tends to last between 0 and 60 seconds. This approach allows us to capture the majority of continuous traffic rate motif patterns.

\noindent{\bf Traffic Rates to Image Representation}. Eventually, we covert time-series network traffic rates into four different type of image representation formats. We  first segment traffic rate data. Each segment is transformed into RGB images using Line Chart, Heat Map, Scatter Plot, and GAF images. We also drop some segments, since there are barely any traffic motifs presented and it could not meet the minimal account of training data requirement and thus are insufficient to represent traffic characteristics of a device.

\noindent{\bf Ethical Consideration for Data Management}. We use open-source datasets to explore the severity and extent of user privacy threat from IoT device traffic traces. We did not collect data from people in this project. Our long-term goal is to provide system solutions to enable people to regain the control of privacy leakages through their traffic rate data. We removed user identical information and sampled the datasets. We do not plan to release our TA attack model codes to the public. {\emph{We followed our institution's Institutional Review Board (IRB) exempt process}.

\subsection{Experimental Setup} 

\subsubsection{TA Attack Models}

We implement a wide set of TA attack models based on the threat models and attack models from prior works~\cite{Ding:2018:SID:3243734.3243865,park2014energy,cai2014systematic,chen2014combined,bovornkeeratiroj2020repel,wang2017walkie,wang2008dependent,shmatikov2006timing,apthorpe2019keeping,dyer2012peek,juarez2016toward}. We use the following models to comprehensively evaluate the prior work in the literature: 1) ML an DL enabled attacks using Logistic Regression, Decision Tree, Support Vector Machines (SVMs), Random Forest, and CNNs; 2) our new image-based attack model using four image representations.

\subsubsection{TA Attack Defense Models}

Regarding TA defense approaches, we will the following TA defense models to evaluate the performance of prior works through ITEMTK.

\noindent{\bf Pure Traffic Injection (PTI)}. We first implement a general version of prior work~\cite{park2014energy,cai2014systematic,liu2018epic}. This approach leverages Bernoulli distribution and Poisson distribution to randomly inject artificial traffic spikes that are randomly selected from historical traffic motifs.

\noindent{\bf Random Traffic Padding (RTP)}. We implement a general version of prior work~\cite{apthorpe2019keeping,dyer2012peek,juarez2016toward}. This approach employs a threshold-based traffic rate flattening, and leverage  Bernoulli distribution, Poisson distribution, and Linear Chain Conditional Random Field (LCCRF) to randomly inject ``fake" traffic spikes.

\noindent{\bf Hybrid Traffic Reshaping (HTR)}. We implement a general version of prior work~\cite{chen2014combined,bovornkeeratiroj2020repel,shmatikov2006timing,wang2008dependent,raghavan2013coupled,raghavan2014modeling,hasan2008reconfigurable,rothmeier2020prediction,PrivacyGuard}. This approach employs traffic rate flattening, and leverages Hidden Markov Model (HMM)-based user behavior modeling to inject artificial traffic motifs that are randomly selected from historical traffic patterns.

\noindent{\bf PrivacyGuard}. PrivacyGuard~\cite{PrivacyGuard} employs intelligent DCGANs-based IoT device traffic signature learning, Long short-term memory (LSTM)-based artificial traffic signature injection, and partial traffic reshaping to further obfuscate private information that can be externally observed in IoT traffic rate traces. We download and adapt the source code of PrivacyGuard~\cite{PrivacyGuard} into ITEMTK framework.

\noindent{\bf PAROS}. PAROS~\cite{Paros} is another variant of PrivacyGuard~\cite{PrivacyGuard}. For TA attack perspective, the major difference between PrivacyGuard and PAROS is the efficiency of their user behavior model. We also download and integrate PAROS~\cite{Paros} into ITEMTK framework. 

\subsection{Evaluation Metrics}

\noindent{\bf Precision}. The traffic motifs related to user activity showcase varying frequencies. In the context of imbalanced datasets, Precision is a critical measure in such scenarios~\cite{buckland1994relationship}. It assesses the proportion of positive identifications that were actually correct. Precision values range from $0.0$ to $1.0$, where $1.0$ signifies that every instance predicted as positive is indeed positive, while $0.0$ means no predicted positives were correct. The formula for computing Precision is given as follows, where TP represents the number of true positives and FP represents the number of false positives:

\begin{equation}
  \text{Precision} = {TP}/({TP + FP})  
\end{equation}

\noindent{\bf Recall}. Recall measures the proportion of actual positives that are correctly identified by the classifier. It is crucial in situations where the cost of missing a positive instance is high. The value of Recall ranges from 0.0 to 1.0, with 1.0 representing a model that correctly identifies all positive instances, and 0.0 indicating that the model fails to identify any positive instances. The expression for Recall:

\begin{equation}
    \text{Recall} = {TP}/({TP + FN})
\end{equation}

In essence, recall is about capturing as many positives as possible, while precision is about being correct in the positive predictions made. Balancing these two measures is often necessary because improving recall typically reduces precision and vice versa.

\noindent{\bf F1 Score}. To quantify the accuracy of TA attack and defense methods, we plan to use F1 score~\cite{huang2015maximum}, which can be defined as a harmonic mean of the precision and recall, where an F1 score reaches its best object or motion detection accuracy at 1 and worst score at 0. The relative contribution of precision and recall to the F1 score are equal. In the multi-class and multi-label case, this is the average of the F1 score of each class with weighting depending on the average parameter. F1 score can be defined as follows, 

\begin{equation}
F1 = 2 * (precision * recall) / (precision + recall)
\end{equation}

\noindent{\bf Matthews Correlation Coefficient (MCC)}. We use the MCC~\cite{mcc}, a standard measure of a classifier's performance, where values are in the range $-1.0$ to $1.0$, with $1.0$ being perfect object detection, $0.0$ being random object inference, and $-1.0$ indicating object inference is always wrong, to benchmark different approaches. The expression for computing MCC is below, where TP is the fraction of true positives, FP is the fraction of false positives, TN is the fraction of true negatives, and FN is the fraction of false negatives. 

\vspace{-0.25cm}
\begin{equation}
\frac{TP*TN - FP*FN}{\sqrt{(TP+FP)(TP+FN)(TN+FP)(TN+FN)}}
\end{equation}

\noindent{\bf Top-k Accuracy}. Top-k accuracy means that the correct class gets to be in the Top-k probabilities for it to count as ``correct''.  This metric computes the number of times where the correct label is among the Top-k labels predicted (ranked by predicted scores).

\subsection{Experimental Results}

\begin{table}[t!]
\small
\begin{center}
\begin{tabular}{||c||c|c|c|c||c||}
\hline
\textbf{Models} & \textbf{Defenses} & \textbf{Precision} & \textbf{Recall} & \textbf{\makecell{F1\\ Score}} & \textbf{MCC} \\ \hline
\multirow{6}{*}{\makecell{Random\\ Forest}} & {Original} &0.516 &0.500 &0.410 &0.388 \\
\cline{2-6} 
~ & {PTI}  & 0.414 & 0.468 & 0.370 & 0.346 \\
\cline{2-6} 
~ &{RTP} & 0.458 & 0.449 & 0.360 & 0.309 \\ 
\cline{2-6} 
~ & {HTR}  & 0.406 & 0.452 & 0.356 & 0.314 \\ 
\cline{2-6} 
~ & {PrivacyGuard}  & 0.41 & 0.472 & 0.386 & 0.343 \\ 
\cline{2-6} 
~ & {PAROS} & 0.449 & 0.496  & 0.417& 0.377  \\ \hline
\multirow{6}{*}{\makecell{Logistic\\ Regression}} & {Original}  &  0.294&  0.322 & 0.269 & 0.146 \\
\cline{2-6} 
~ & {PTI}  &0.210  & 0.268 & 0.198 & 0.060 \\
\cline{2-6} 
~ &{RTP} &0.234  &0.320  &0.208  &0.095 \\ 
\cline{2-6} 
~ & {HTR} & 0.187 & 0.235 & 0.197 &0.046  \\ 
\cline{2-6} 
~ & {PrivacyGuard}  &0.189  &0.241  &0.199  & 0.053  \\ 
\cline{2-6} 
~ & {PAROS}  & 0.229 &0.292  &0.200 &0.071 \\ \hline

\multirow{6}{*}{\makecell{Decision\\ Tree}} & {Original}  &0.366 &0.374 &0.370 &0.254 \\
\cline{2-6} 
~ & {PTI}  &0.345  & 0.349 & 0.346 &0.224  \\
\cline{2-6} 
~ &{RTP}  &0.324  &0.323  &0.323  &0.196
\\ 
\cline{2-6} 
~ & {HTR}  & 0.306& 0.317 & 0.311 &0.181  \\ 
\cline{2-6} 
~ & {PrivacyGuard}  &0.327  &0.330  &0.328  &0.204  \\ 
\cline{2-6} 
~ & {PAROS} & 0.350 & 0.356 &0.352 & 0.230\\ \hline

\multirow{6}{*}{CNN} & {Original}  &0.261 &0.345 &0.194 &0.135 \\
\cline{2-6} 
~ & {PTI}  &0.103  & 0.321 & 0.156 &0.067  \\
\cline{2-6} 
~ &{RTP}  &0.177  &0.329  &0.173  &0.075 \\ 
\cline{2-6} 
~ & {HTR}  &0.131  &0.319  &0.159  &0.011  \\ 
\cline{2-6} 
~ & {PrivacyGuard}  &0.174  &0.320  &0.165  &0.033  \\ 
\cline{2-6} 
~ & {PAROS}  & 0.133 & 0.319 & 0.161& 0.018\\ \hline
\multirow{6}{*}{\makecell{Image\\ Attack}} & {Original}  &0.794 &0.790 & {\textbf{0.781}} & {\textbf{0.722}} \\
\cline{2-6} 
~ & {PTI}  &0.596  & 0.589 & {\textbf{0.592}} &{\textbf{0.415}}  \\
\cline{2-6} 
~ &{RTP}  &0.678  &0.695  &{\textbf{0.687}}  &{\textbf{0.472}} \\ 
\cline{2-6} 
~ & {HTR}  &0.517  &0.532  &{\textbf{0.524}}  &{\textbf{0.360}}  \\ 
\cline{2-6} 
~ & {PrivacyGuard}  &0.481  &0.531  &{\textbf{0.493}}  &{\textbf{0.496}}  \\ 
\cline{2-6} 
~ & {PAROS}  & 0.626 & 0.642 & \textbf{0.620}& \textbf{0.473}\\ \hline
\end{tabular}
\end{center}
\caption{The performance of TA defense approaches when encountering different TA attacks.}
\label{table:attack-models-all}
\vspace{-0.3in}
\end{table}

\subsubsection{Preventing User Activities Detection Attacks}
\label{sec:user-activities-attacks}

We first benchmark the effectiveness of hiding user activities when applying five different TA attack preventing approaches. We leverage ML/DL-based attack models that we built in Section~\ref{sec:design} to detect 14 different user activities using the original, PTI modified, RTP modified, HTR modified, PrivacyGuard, and PAROS modified traffic rate traces. Unsurprisingly, as shown in Table~\ref{table:attack-models-all}, all the F1 scores and MCCs are dropping after users deploying the most robust preventative measures. We find that all the TA defense approaches are indeed safeguarding smart homes. Interestingly, all the defense approaches yield significantly lower MCCs when encountering Logistic Regression and CNNs powered TA attacks. While, all the TA defense approaches cannot sufficiently protect user privacy when handling Decision Tree and Random Forest based TA attacks. Current TA defenses can only hide user privacy well under certain attacks.

\begin{figure*}[t!]
\begin{center}
\includegraphics[width=0.95\textwidth]{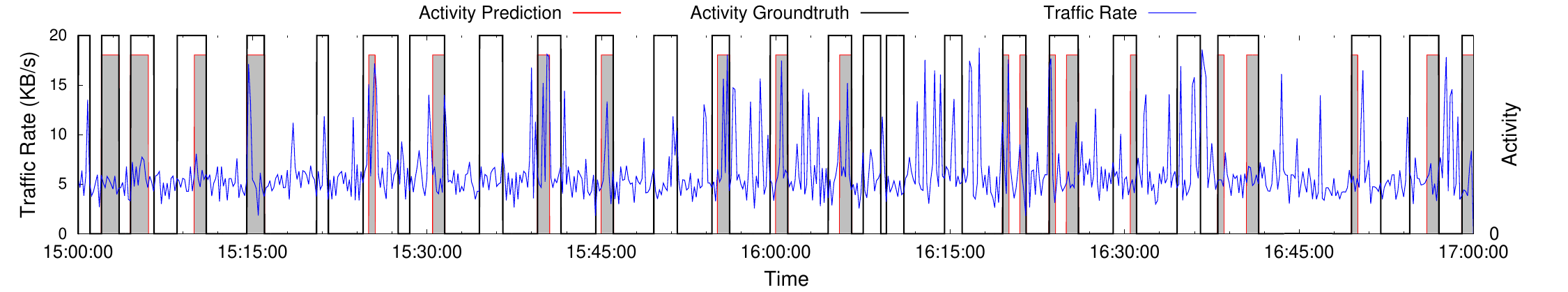}
\end{center}
\vspace{-0.2in}
\caption{User activity attack results on the defenses reshaped data.}
\label{fig:activity_prediction}
\vspace{-0.2in}
\end{figure*}

In addition, we find that our new image based attack yields the best F1 score and MCC across \textit{all} the recent defense works---PTI, RTP, HTP, PrivacyGuard, and PAROS. It shows that our novel image-based attack is capable of inferring sensitive user information, even when users employing the most robust preventative measures in their smart homes. Figure~\ref{fig:activity_prediction} illustrates the performance of ITEMTK's image-based TA attack. Y1 axis is traffic rate, which is measured in KB per second. Y2 axis is the user activity signal that indicates ITEMTK identifies at least one user activity. The gray areas in each bar represents the groundtruth user activities that had been detected. Even using recent TA attack defense approaches, our image attack can still significantly reveal groundtruth user activities. This is mainly due to the fact that there are still significant residual genuine traffic patterns left in the traffic rates which are already reshaped by the most recent TA defense approaches. Thus, our results show that current defending approaches are not sufficient to protect IoT device user privacy and thus smart home IoT devices are significantly vulnerable to our new image-based user privacy inference attacks, posing a grave threat to IoT privacy.

\begin{figure}[t!]
\begin{center}
\includegraphics[width=0.35\textwidth]{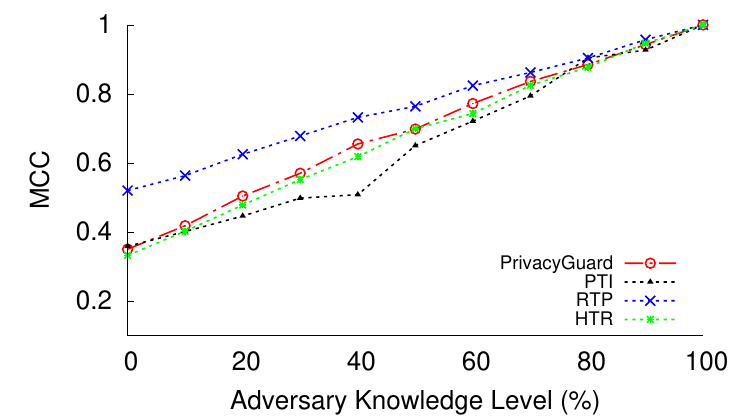}
\end{center}
\vspace{-0.3cm}
\caption{ITEMTK's performance under adaptive adversaries.}
\label{fig:ac}
\vspace{-0.2in}
\end{figure}

\noindent{\bf Results}: \emph{Although the most recent TA attack defense approaches, including PTI, RTP, HTP, PrivacyGuard and PAROS, can efficiently prevent some Logistic Regression and CNNs powered TA attacks, they are all significantly vulnerable to our new image representation-based user privacy inference attacks, posing a grave threat to user privacy}.

\subsubsection{Preventing User Activities Detection by Adaptive Adversary}

We next examine the effect of adversary confidence (AC) on ITEMTK's performance. Figure~\ref{fig:ac} shows the ability of ITEMTK to preserve user privacy when adaptive adversary having more knowledge about our TA attack and defense model. The attacker knowledge level is described as the percentage of traffic rate testing dataset that an external adversary poses to train the adversarial ML/DL-based models to infer user activities. As shown in Figure~\ref{fig:ac}, 0\% indicates that an external adversary has no knowledge of the targeted home testing dataset and thus no cross-validation is performed in their modeling, while, 100\% means that the external adversary has observed all the groundtruth traffic patterns for each user activity such that the attack models are ``perfectly'' trained and tested using the same testing dataset. The goal is to understand the ability of the TA defense approaches when protecting user privacy under the attacks from the ``adaptive'' adversaries, who have different knowledge levels about traffic rates of the target smart home IoT devices. Our findings in Figure~\ref{fig:ac} reveal that as the attacker's knowledge level escalates, the TA attack defense methods within ITEMTK demonstrate a higher MCC. The attack accuracy, measured by MCC, exhibits a linear correlation with the attacker's level of knowledge.

\noindent{\bf Results}: \emph{The most recent TA attack preventing approach in ITEMTK exhibits a linear correlation with the attacker's level of knowledge about a smart home. Thus, current defenses cannot sufficiently protect user privacy under attacker's dynamic knowledge level scenarios.}

\begin{table}[t!]
\small
\begin{center}
\begin{tabular}{||c||c|c||c||c|}
\hline
\textbf{Countermeasures} & \textbf{Top 1} & \textbf{Top 5} & \textbf{{MCC}} & \textbf{{F1}} \\ \hline
{PTI} & 75.3\% & 91.1\% &0.705 &0.743  \\ \hline
{RTP} & 86.5\% & 97.9\%  & 0.868 &0.864 \\ \hline
{HTR} &77.4\% & 95.5\% & 0.767 &0.775  \\ \hline
{PrivacyGuard~\cite{PrivacyGuard}} & 78.6\% & 94.7\% & 0.774 & 0.784  \\ \hline
{PAROS~\cite{10230103}} & 80.1\% & 94.6\%  & 0.787 &0.809  \\ \hline
\end{tabular}
\end{center}
\caption{The comparison of Image-based device type attack when applying different defending approaches.}
\vspace{-0.4in}
\label{table:device-type-}
\end{table}

\subsubsection{Quantifying IoT Device Inference Attack Prevention (Attack Scenario \#1)} Next, we quantify the effectiveness of preventing IoT device detection when applying five different TA attack preventing approaches. In Section~\ref{sec:user-activities-attacks}, we find that although the most recent TA attack defense approaches can efficiently prevent some Logistic Regression and CNNs powered TA attacks, they are all still significantly vulnerable to our new image representation-based user privacy inference attacks. Table~\ref{table:device-type-} illustrates the results of our new image-based device type attack after applying different recent TA attack defending approaches. Our results show that our image-based new attack within ITEMTK yields extremely elevated MCCs ($\geq$ 0.71) and F1 scores ($\geq$ 0.74) across different defending approaches, including PTI, RTP, HTR, PrivacyGuard and PAROS. Meanwhile, we're witnessing an exceptionally high Top-5 accuracy ($\geq$ 91\%). This is mainly because our image-based new attack could examine the residual genuine traffic motifs at different granularities of IoT traffic rates using the image fusion network simultaneously. Thus, current defending approaches cannot sufficiently prevent our image-based device type attack, even when users employ the robust preventative measures, posing a grave threat to user privacy.

\noindent{\bf Results}: \emph{Image-based attack from ITEMTK yields very high MCC and F1 score when inferring IoT device type, even when users employ the most robust preventative measures. ITEMTK has shown that current defenses cannot sufficiently prevent our new image-based device type attack, presenting a serious threat to the privacy of IoT device users.}

\subsubsection{Quantifying IoT Device Event Inference Attack Prevention} Similar, we next plan to quantify the effectiveness of preventing IoT device event detection when applying five different TA attack preventing approaches. By doing so, we are benchmarking effectiveness and completeness of most recent TA attack preventing approaches on ``micro'' use sensitive information. One user activity may have multiple IoT device events involved and thus present more comprehensive or longer term user personal information leakage. Instead, IoT device event is leaking more detailed user in-home behaviors. Table~\ref{table:device-event-} illustrates the attacking accuracy results when when applying five different TA attack preventing approaches on a smart home's network traffic rates. Our results show that even when users employ the most robust preventative measures in their smart homes, our image-based attack model can still achieve very high MCCs ($\geq$ 0.41) and F1 Scores ($\geq$ 0.65). We are also witnessing an exceptionally high Top-5 accuracy ($\geq$  93\%). Thus, most recent TA defense approaches could not successfully and sufficiently mask IoT device event information, which embedded in their traffic rates.

\noindent{\bf Results}: \emph{Image-based attack from ITEMTK yields very high MCC and F1 score when inferring IoT device events, even when users employ the most robust preventative measures. ITEMTK has shown that current defenses cannot sufficiently prevent our image-based device event attack, presenting another serious threat to smart home user privacy.}

\begin{table}[t!]
\small
\begin{center}
\begin{tabular}{||c||c|c||c||c|}
\hline
\textbf{Countermeasures} & \textbf{Top 1} & \textbf{Top 5} & \textbf{{MCC}} & \textbf{F1}\\ \hline
{PTI} & 81.3\% & 94.8\% &0.622 &0.807  \\ \hline
{RTP} & 73.1\% & 93.5\% &0.503 &0.731 \\ \hline
{HTR} & 66.6\%& 92.7\% &0.433  &0.650 \\ \hline
{PrivacyGuard~\cite{PrivacyGuard}} & 68.1\% & 93.0\% & 0.414 &0.661  \\ \hline
{PAROS~\cite{10230103}} &  69.5\%& 92.7\%  & 0.473 & 0.684\\ \hline
\end{tabular}
\end{center}
\caption{The comparison of Image-based device event inference attack when applying different defending approaches.}
\vspace{-0.2in}
\label{table:device-event-}
\end{table}

\subsubsection{Preventing User Activities Detection Attacks from Single Direction Traffic} Next, we examine the effectiveness of preventing user activity detection when applying five different TA attack preventing approaches on single-direction traffic flows. \emph{The on-path adversaries do not have the capability to monitor the bidirectional traffic flows from a smart home. Instead, they can only sniff either the outgoing or incoming traffic rates. In doing so, we are assessing the severity of the privacy threat posed by our image-based TA attack on single-direction traffic rates.} Table~\ref{table:device-event-discussion} demonstrates the attacking accuracy results when applying five recent TA attack preventing approaches on single-direction network traffic rates. Interestingly, we find that using only incoming traffic rates, our image-based attack model yields the best MCC of $\sim$0.267 and F1 as of $\sim$0.469. Similarly, we observe that our image-based attack model yields the best MCC of $\sim$0.299 and F1 score as of $\sim$0.590 using only outgoing traffic rates. The outgoing traffic rates are more vulnerable to our new image-based inference attacks. This is mainly due to the fact outgoing traffic of IoT devices often are interacting with users in smart homes. For instance, people may talk to their voice assistant, or trigger motion detection sensors in their home. The user interaction activities will be recorded in their outgoing traffic rates. While, incoming traffic are often the responses of users' outgoing requests.

\begin{table}[t!]
\small
\begin{center}
\begin{tabular}{||c||c|c|c|c||c||}
\hline
\textbf{Models} & \textbf{Defenses} & \textbf{Precision} & \textbf{Recall} & \textbf{\makecell{F1\\ Score}} & \textbf{MCC} \\ \hline
\multirow{6}{*}{\makecell{In Data}} 
~ & {PTI}   & 0.457& 0.474 & 0.423 & 0.246\\
\cline{2-6} 
~ &{RTP}  & 0.444 & 0.424 & 0.410 & 0.267 \\ 
\cline{2-6} 
~ & {HTR}  & 0.348 & 0.323 & 0.296 & 0.172 \\ 
\cline{2-6} 
~ & {PrivacyGuard}  & 0.350 & 0.265 & 0.303 & 0.135 \\ 
\cline{2-6} 
~ & {PAROS} & 0.466 & 0.435  & 0.469& 0.208  \\ \hline
\multirow{6}{*}{\makecell{Out Data}}
~ & {PTI}  &0.517  & 0.490 & 0.515 & 0.295 \\
\cline{2-6} 
~ &{RTP} &0.568  &0.590  &0.590  &0.299 \\ 
\cline{2-6} 
~ & {HTR} & 0.477 & 0.426 & 0.469 &0.208  \\ 
\cline{2-6} 
~ & {PrivacyGuard}  &0.462  &0.461  &0.467  & 0.187  \\ 
\cline{2-6} 
~ & {PAROS}  & 0.509 &0.547  &0.525 &0.244 \\ \hline
\end{tabular}
\end{center}
\caption{The image-based attack using single-direction traffic.}
\label{table:device-event-discussion}
\vspace{-0.3in}
\end{table}

\noindent{\bf Results}: \emph{Image-based attack can still yield the MCC as of $sim$0.299 and F1 score as of $sim$0.590 on single-direction traffic rates. The outgoing traffics are more vulnerable to image-based TA inference attacks.}

\subsubsection{System Scalability and Cost Analytics}

We next examine system scalability and cost of ITEMTK framework. Figure~\ref{fig:overhead} demonstrates the system overhead of ITEMTK when our image fusion network is processing different amount of image representations. For each cluster group, we report CPU utilization, CPU RAM usage, GPU RAM usage, and turnaround time for (re)training. In Figure~\ref{fig:overhead}, we demonstrated four clustered groups, including 1, 2, 3, and 4 different image representations, respectively. Our results show that ITEMTK's system overhead across different cluster groups have only a minimal or marginal increase on CPU and RAM utilization. While we do notice a slight increase in GPU usage with an escalation in the number of image presentations, the GPU RAM usage for (re)training the framework remains below 35\%, and the turnaround time remains under approximately 330 seconds per epoch. Note that, depending on real-world applications, ITEMTK might not necessitate retraining of this network. This (re)training process may require 10$\sim$30 epochs and is also up to user personal preference.

\noindent{\bf Results}: \emph{ITEMTK demonstrates the performance consistency, with only a minimal or marginal increase on system overhead in terms of CPU and RAM utilization across four different image fusion networks. In addition, GPU RAM usage for (re)training the framework remains below 35\%, and the turnaround time remains under approximately 330 seconds. Our framework---ITEMTK is able to provide TA attack and defense evaluation with a reasonable and consistent system overhead.}

\subsection{Insights and Future Defense Directions}

\noindent{\bf Residual Genuine Traffic Patterns}. As we discussed in Section~\ref{sec:background} and Section~\ref{sec:design}, our new image-based TA attack is built on top of the ``residual'' genuine traffic patterns that have already embedded user activity information. In particular, this kind of residual genuine traffic patterns might be presented at different granularities of traffic rates. Prior approaches typically on focus on TA attacking or defending on one granularity traffic rate data. Our image-based attack leverages multiple image presentation fusion network to detect  residual hidden genuine traffic patterns. Specially, GAF image presentation could observe hidden genuine traffic patterns from traffic rates at different granularities. To prevent image-based TA attacks, it would be helpful if we could build new TA attack defense approaches at different traffic rate granularities. Masking the remaining genuine traffic patterns at different traffic rate granularities simultaneously could help prevent image-based TA attacks.

\begin{figure}[t!]
\begin{center}
\includegraphics[width=0.4\textwidth]{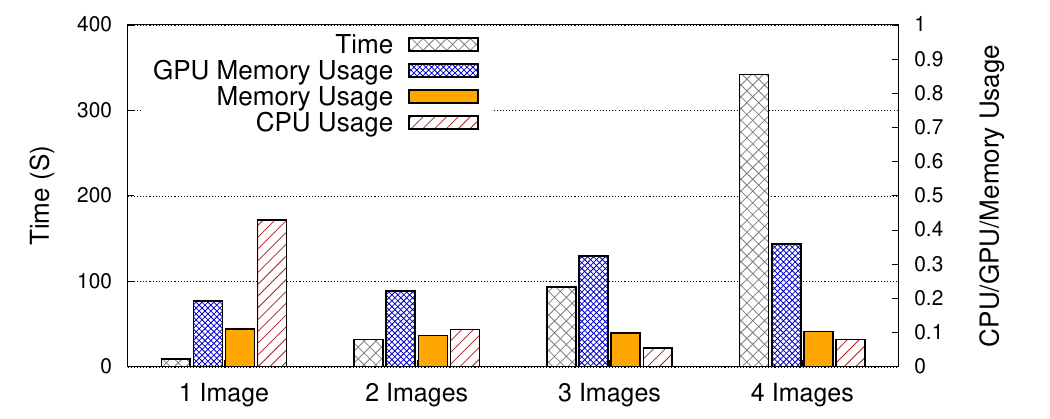}
\end{center}
\vspace{-0.3cm}
\caption{System Overhead Analytics.}
\label{fig:overhead}
\vspace{-0.2in}
\end{figure}

\noindent{\bf Absolute vs Relative Traffic Reshaping}. Existing methods for defending against traffic analysis utilize either static or dynamic thresholds to conduct partial reshaping of traffic rates and inject traffic rates guided by ML/DL user behavior models. These reshaping techniques often change the ``spikes'' directly presented in traffic rates. However, as shown in Figure~\ref{fig:summary} and Figure~\ref{fig:GAF-number}, when converting these residual traffic rates into various image representations, it is still possible to visualize the correlations before and after the application of the latest reshaping approach—PrivacyGuard. This is primarily because, even though the absolute time-series traffic rate data has been altered, the relative traffic patterns depicted in their transformed image representations (partially) persist. Thus, our image-based TA attack could leverage this information to infer user activities. To mitigate the risk of image-based TA attacks, we could modify the traffic patterns in their image representations, in addition to reshaping the original time-series traffic patterns.

\noindent{\bf Artificial Device Injection}. 
Another fundamental factor leading to privacy leakage is the occurrence of IoT device events. An effective approach involves injecting and replaying traffic patterns from an artificial device. This can be elucidated by homeowners acquiring new IoT devices. For example, in the context of an Amazon Alexa-connected smart home, we could replay a series of Google Home traffic patterns. Using the same dataset outlined in Table~\ref{table:attack-models-all}, our image-based attack model demonstrates the F1 score of 0.152 and MCC of 0.121. These values are three times lower than those achieved by PAROS, as indicated in the same table.

%% file: conclusion.tex
We design a new low-cost, open-source systematic framework---ITEMTK that enables people to comprehensively examine and validate prior traffic analytics (TA) attack models and their defending approaches. Specially, we also design a novel image-based TA attack that could infer sensitive user information, even after users deploying the most recent robust TA attack preventative approaches in their smart homes. Our results show that current defending approaches cannot sufficient protect user privacy. IoT devices are significantly vulnerable to our new image-based user privacy inference attacks, posing a grave threat to IoT device user privacy. We also highlight potential future improvements to enhance the current TA attack defending approaches using initial results. ITEMTK is a versatile toolkit that enables users to easily expand its functionality by integrating new TA attacks and prevention approaches, providing a benchmark for their future work. We plan to use more datasets to further benchmark the performance of image-based TA attack and the utility of ITEMTK framework. We will investigate additional image representation approaches, which could potentially provide more efficient image fusion functionality for ITEMTK. We will also design a new traffic reshaping approach that could significantly prevent image-based TA attacks on IoT traffic rates. 

\noindent{\textbf{Acknowledgements}}. This research is supported by NSF grant CNS-2238701.